\documentclass[12pt,showkeys,altaffilsymbol]{revtex4}
\usepackage{graphicx}   

\newcommand{\Vbg}{$V_{\rm bg}$ }
\newcommand{\Vtg}{$V_{\rm tg}$ }
\newcommand{\Cbg}{C_{\rm bg}}
\newcommand{\Ctg}{C_{\rm tg}}
\newcommand{\Vds}{$V_{\rm ds}$ }
\newcommand{\Vj}{$V_{\rm j}$ }
\newcommand{\dVdI}{$dV_{\rm ds}/dI$ }
\newcommand{\pn}{$p$-$n$ }

\newcommand{\AlOx}{AlO$_x$ }
\newcommand{\figwidth}{0.9\columnwidth}
\newcommand{\MANA}{MANA, National Institute for Materials Science, Namiki, Tsukuba 305-0047, Japan}
\newcommand{\CREST}{CREST, Japan Science and Technology Agency, Kawaguchi 332-0012, Japan}
\newcommand{\NEC}{NEC Corporation, Tsukuba 305-8501, Japan}
\newcommand{\UT}{Institute of Physics, University of Tsukuba, Tsukuba 305-8571, Japan}

\begin{document}
\title{Observation of Tunneling Current in Semiconducting Graphene \pn Junctions}

\author{Hisao Miyazaki} \altaffiliation{\CREST} \affiliation{\MANA}
\author{Michael Lee} \affiliation{\MANA}
\author{Song-Lin Li} \affiliation{\MANA}
\author{Hidefumi Hiura} \altaffiliation{\NEC} \affiliation{\MANA} 
\author{Kazuhito Tsukagoshi} \altaffiliation{\CREST} \email{TSUKAGOSHI.Kazuhito@nims.go.jp}  \affiliation{\MANA}  
\author{Akinobu Kanda} \altaffiliation{\CREST} \affiliation{\UT}

\date{\today}


\begin{abstract}
We demonstrate a tunneling and rectification behavior in bilayer graphene. A stepped dielectric top gate creates a spatially modulated electric field, which opens the band gap in the graphene and produces an insulating region at the \pn interface. A current-voltage relationship exhibiting differential resistance peak at forward bias stems from the tunneling current through the insulating region at the \pn interface. The tunneling current reflects singularities in the density of states modified by the electric field. This work suggests that the effect of carrier charge tuning by external electric field in 2D semiconductors is analogously to that by impurity doping in 3D semiconductors.
\end{abstract}

\keywords{bilayer graphene, field effect, $p$-$n$ junction, tunneling effect}

\maketitle


\section{Introduction}
In an atomically-thin two-dimensional (2D) conductor like graphene, carrier charges can be induced with an external electric field\cite{novoselov}. Inducing carrier charges by an electric field is one significant advantage of 2D conductors, because it provides uniformly distributed carrier charges and overcomes the problem of random spatial fluctuation of dopants\cite{Intel45nm}. This random fluctuation is one of the important problems in nanoelectronics that use traditional intrinsic semiconductors. A top gate that induces distinct localized electric fields in the 2D conductor can produce distinct regions with different carriers, thus forming a \pn junction in graphene that are required for devices\cite{HuardPRLpnp, GorbachevNLpnp}. Klein tunneling\cite{AndreaNPhysKlein, StanderPRL102Klein} and quantum Hall edge modes\cite{WilliamsSciencePNQHE, OzyilmazPRL99pnp} have been observed in these junctions. In monolayer graphene, such gate-controlled \pn junctions have been demonstrated increase the resistance through the active regions. However, an efficient barrier to current flow that will produce diode current-voltage characteristics has not yet been reported. An efficient barrier requires a potential barrier region to be induced in a 2D conductor with a bandgap. 

In this paper, we report the realization of such a barrier region at a \pn junction in bilayer graphene (BLG), the first observation of tunneling through a \pn junction induced by electric fields in semiconducting BLG, and the first instance of rectification in a device based on a 2D material. We observed a differential resistance peak at forward source-drain bias. We attribute this peak to tunneling between $p$ and $n$ regions, and credit it for diode behavior similar to what is observed in a 3D Esaki diode\cite{sze,Esaki1958, EsakiJPSJ}. We employed a uniform bottom gate and a stepwise top gate to form a \pn interface. A thin top gate dielectric (less than 10 nm) \cite{miyazakiSST} is critical to shrink the spatial transition between $p$ and $n$ regions. The electric field between the top gate and the bottom gate opens a band gap in the BLG\cite{mccann, JPSJ78AndoKoshinoBLG, oostinga, zhang, mak, MiyazakiNLbilayergap}, making a tunnel barrier between $p$ and $n$ regions. As this is the first instance of rectification in 2D, we anticipate our approach will provide a starting point for creating gate-controlled diodes in 2D conductors. Furthermore, a gate-controlled \pn junction would be utilized in optoelectronic devices operating in the THz regime\cite{THzgap, RyzhiiPRBTHz}, which is covered by the gate-tunable range of the band gap in BLG. Utilizing the tunneling effect has an advantage for high speed electronics which would be one major application area for graphene electronics\cite{MericProcIEEE}.

\section{Experimental Approach}
An electric field produced by a stepped-gate can generate opposing charges under each half of the gate (Fig.\ 1(a)). A charge-neutral region exists under the step. If the 2D semiconductor has a band gap, charge carriers are depleted in the charge-neutral region. 
The depletion region has an in-plane electric field which originates from the gate electric field, canceled partially by the electric field from charge carriers in the 2D semiconductor. This mechanism sharply contrasts the impurity-doped \pn junction in a 3D semiconductor, where carrier recombination produces a depletion region. The depletion of carriers leaves charged donor and acceptor impurities, resulting in a built-in electric field\cite{sze}. We note that the charge-neutral region and charged region in the gate-controlled \pn junction is opposite from the impurity-doped one: 
In the gate-controlled one, the depletion region is charge-neutral, while $p$ and $n$ regions are charged by carriers. In the impurity-doped one, the depletion region is charged by dopants (donors and acceptors), and uniform regions are charge-neutral. In spite of these differences, a model based on electrostatics suggests that the gate-controlled \pn junction in 2D mirrors the operation of the impurity-doped one in 3D (Appendix A).

\section{Sample Fabrication}
BLG samples were prepared from kish graphite by the mechanical cleavage method\cite{novoselov} and adhered on to a highly-doped Si substrate with a 90-nm-thick SiO$_2$ surface layer. A graphene sample with multiple electrodes was patterned by oxygen plasma etching for four-terminal measurements (Fig. 1(b)). The bilayer channel was 0.4 $\mu$m in width and was sandwiched between a substrate bottom gate and a stepped top gate. The top gate was composed of two regions with different gate dielectric thicknesses. Half of the area of the graphene channel (surrounded by dashed lines in Fig.\ 1(b)) was covered by a 5-nm-thick layer of SiO$_2$. Then, the entire area of the graphene between the voltage terminals was covered by a 30-nm-thick Al film. The sample was exposed to air for several hours for partial oxidization of the Al film. An oxidized (AlO$_x$) layer formed not only on the surface but also at the interfaces of Al/graphene\cite{MiyazakiNLbilayergap, miyazakiAPEXscr, yiAlOx} and Al/SiO$_2$\cite{bauer:1006, ReacAlSiO2}. The SiO$_2$ layer between part of the graphene and the \AlOx layer, increases the dielectric thickness over that region\cite{LiSmall} which results in the formation of a stepwise junction in the top gate (Fig.\ 1(c)). When a voltage is applied to the top gate, two different electric fields are simultaneously applied to the individual regions. The field effect mobilities extracted from the gate voltage dependence were 1300 cm$^{-2}$V$^{-1}$sec$^{-1}$ for electrons and 1800 cm$^{-2}$V$^{-1}$sec$^{-1}$ for holes, independent of the dielectric thickness. 

\section{Results and Discussion}
The two gate regions with different dielectrics thicknesses create charge neutrality point (CNP) ridges with different slopes in the $V_{\rm bg}$-$V_{\rm tg}$ plane. Figure 1(d) shows a contour plot of the resistance $R_0$ through the junction as a function of \Vbg and $V_{\rm tg}$. 
The two ridges separate the $V_{\rm bg}$-$V_{\rm tg}$ plane into $p$-$p$, $p$-$n$, $n$-$p$, or $n$-$n$ combinations of carrier polarity in the graphene. 
On the ridge of CNPs, the height of the resistance peak increased with an increasingly negative electric field ($V_{\rm bg}\to -40$ V) as an evidence of the band gap opening under the electric field\cite{MiyazakiNLbilayergap}. 
From the slope of the CNP lines, we can extract the top gate capacitance values\cite{HuardPRLpnp}: $C_{\rm tg1}=9.3\times 10^{-3}$  F/m$^2$ in the thick region and $C_{\rm tg2}=5.5\times 10^{-3}$  F/m$^2$ in the thin region. 
If we assume the dielectric film is simply composed of SiO$_2$, equivalent thicknesses are approximately 3.7 nm ($= d_1$) and 6.3 nm ($= d_2$), respectively.  The stepwise top gate causes a transient region of $\sim 5$ nm which gives a width of the insulating region $W_{\rm D}$ (Fig.\ 1(e) and Appendix A).

In an actual bilayer of graphene with a band gap, a disordered potential forms a band tail\cite{ElliottAmorphous}. The band structure forms two peaks in the density of states (DOS) at the edge of the valence band and the conduction band\cite{nilsson,mkhiraryan}.
The energy gap between the two peaks $E_{\rm g}$ hardly depends on disorder, and is almost identical to the band gap in the unperturbed system\cite{MinOptgap}. This gap
can be extracted from the temperature dependence of the conductance at the CNP\cite{MiyazakiNLbilayergap}.
The CNP conductance comprises the intrinsic band conduction and the hopping conduction via localized states which make up the band tail.
The former has a thermal activation energy $E_{\rm g}/2$.
Extracted $E_{\rm g}$ is $\sim 0.2$ eV at $V_{\rm bg} = -40$ V both in the thin and thick dielectric region. The band gap makes an insulating  region inserted between the $p$ and $n$ regions. 
Using this band gap, we estimate tunneling probability across the junction to be 0.07 and the tunnel resistance $R_{\rm t}$ to be several k$\Omega$ (Appendix B). 
The hopping conduction causes leakage current coexisting with the tunnel conduction. The leakage resistance is estimated to be $R_{\rm L} \sim 1$ k$\Omega$ (Appendix B), which is comparable to the tunnel resistance. Then, the junction resistance, $R_{\rm j} = 1/(1/R_{\rm t} + 1/R_{\rm L})$, is on the order of 1 k$\Omega$. Using this value, it is estimated that the voltage drop at the junction \Vj is a few percent of $V_{\rm ds}$.

The differential resistance \dVdI was measured in a four-terminal configuration across the \pn junction as a function of the source-drain voltage $V_{\rm ds}$, rather than the leakage current, to investigate the junction property because the differential resistance is sensitive to a nonlinear tunneling current. 
We found a \dVdI peak for a large bottom gate voltage ($V_{\rm bg} = -40$ V) that opens the band gap (Fig.\ 2(a)).
The peak was not observed for a bottom gate voltage ($V_{\rm bg} = -32$ V) that was too small to open the band gap (bottom-right inset of Fig.\ 2(a)).
The increase of the \dVdI at large \Vds is caused by charge redistribution by the \Vds bias\cite{MericNatNano}.
The peak appeared at a forward bias of $V_{\rm ds}\sim 50$ mV, regardless of \Vbg or \Vtg (Fig.\ 2(b)). 
The peak height depends on the bottom gate voltage and became pronounced when the gate electric field was increased by applying \Vbg (Fig.\ 2(c)).

We analyze the \dVdI peak observed in the experiment. A DOS diagram of the unbiased \pn junction is illustrated in Fig. 3(a)i). The voltage drop, $V{\rm j}$, depresses the energy in the $p$-type side (Fig. 3(a)ii)). The tunneling current, $I_{\rm t}$, from the $p$ to $n$ region is given by $I_{\rm t}\propto T_{\rm t}\int\left[ f(E-E_{\rm F})-f(E-E_{\rm F}+eV_{\rm j})\right] D_n(E)D_p(E+eV_{\rm j})dE$, where $f(x)=1/\left[\exp (x/k_{\rm B}T)+1\right]$ is the Fermi distribution function, and $D_p(\varepsilon)$ and $D_n(\varepsilon)$ are the DOS at the energy level $\varepsilon$ (measured from the mid-gap) for the $p$-type and $n$-type sides\cite{sze}. The tunneling current reaches a maximum when the peaks in $D_p(E + eV_{\rm j})$ and $D_n(E)$ align at the energy level (Fig.\ 3(a)ii)). For a larger $V_{\rm j}$, the tunneling current becomes smaller because the DOS peaks go out of alignment (Fig.\ 3(a)iii)). As a result, the tunneling current has a peak that is a function of $V_{\rm ds}$. In the total current, the tunneling current becomes indistinguishable from the leakage current via localized states when the leakage current is comparable to or larger than the tunneling current. The peak structure displayed a \dVdI peak, as shown in Fig.\ 3(b), which was observed at $V_{\rm ds}\sim 50$ mV. Because the voltage drop at the junction is a few percent of $V_{\rm ds}$, the energy difference between the Fermi level and the band edge must be on the order of a few meV.

The temperature dependence of the \dVdI peak (Fig.\ 4(a)) gives reasonable support to the model of the tunnel junction discussed above. Using the Sommerfeld expansion in terms of temperature\cite{AshcroftMermin}, the tunneling current is proportional to $T_{\rm t}\int_{E_{\rm F}-eV_{\rm j}}^{E_{\rm F}}\left[ P(E)+a_1(k_{\rm B}T)^2 P''(E) \right]dE$  
within the second order of the temperature $T$, where $P(E) = D_n(E)D_p(E + eV_{\rm j})$, $P''(E)=d^2P(E)/dE^2$, and $a_1 = \frac{1}{2} \int_{-\infty}^\infty x^2 \left( -\frac{d}{dx}\frac{1}{\exp(x)+1} \right) dx \sim 1.6$. 
The function $P(E)$ represents a density of states for elastic tunneling between the $p$ and $n$ region.
Because a temperature coefficient for the second order is proportional to the integration of $P''(E)$ around the Fermi level, a trend of the tunneling current in the temperature dependence is determined by convex upward or downward in the $P(E)$.
For a peak in tunneling current, $P(E)$ has a peak around the Fermi level, i.e.\ $P''(E_{\rm F})<0$ (Fig.\ 4(b)).
Thus, the tunneling current decreases with increasing $T$.
This trend can be observed in the temperature dependence of $dV_{\rm ds}/dI$-$V_{\rm ds}$, as shown in Fig.\ 4(a); the decrease in the tunneling current peak causes the \dVdI peak to diminish as the temperature increases (Fig.\ 4(c)).
The decrease in the tunneling current peak also leads to the \Vds dependence in the $dV_{\rm ds}/dI$-$T$ curve (Fig.\ 4(d)). 
At a low \Vds bias ($V_{\rm ds} = 0$), \dVdI becomes larger at higher temperatures, reflecting the decrease in the tunneling current. In contrast, at higher \Vds bias voltages, \dVdI becomes smaller at higher temperatures, reflecting the decrease in the \dVdI peak (Fig.\ 4(c)). 
Here, a second-order temperature coefficient ($1.3\times 10^{-5}$ K$^{-2}$) is extracted from the $dV_{\rm ds}/dI$-$T$ curve at the \dVdI maximum ($V_{\rm ds}=50$ mV). A corresponding energy for the temperature coefficient is 20 meV. Because the temperature coefficient ($\propto P''(E_{\rm F})$) determines the sharpness of the $P(E)$ peak, the corresponding energy represents a broadness of the peak, caused by band tail of the DOS.
This indicates that the band tail width is approximately 20 meV which on the order of 10\% of the energy gap $E_{\rm g}$.

The thermal energy at higher temperatures also generates a leakage current via localized states. The leakage current becomes larger with higher $T$, while the tunneling current becomes smaller. The tunneling current is more dominant than the leakage current in the temperature dependence, because the \dVdI at $V_{\rm ds} = 0$ becomes larger at higher temperature. The overall trend of the \dVdI peak observed in this experiment is similar to the negative differential resistance (NDR) of the Esaki diode\cite{sze, Esaki1958, EsakiJPSJ}. For a larger NDR in graphene, the leakage current must be suppressed. The impurity states in the band gap under high electric field are the most probable cause of the leakage current. Thus, reducing impurities and defects is important for fabricating a diode device governed only by the tunneling effect.

\section{Conclusions}
We observed clear tunneling signals in semiconducting BLG \pn junctions. 
This provides the first experimental evidence for the existence of conduction barrier in gap tunable atomic-layer conductors, which is essential to realize wavelength-tunable optoelectronic devices. 
We also identified localized states as the source of the diffusion current within the band gap, which highlights the importance to exclude the impurities and disorder in graphene to improve performance. With appropriate biasing conditions and transparent top gate stacks for THz electromagnetic waves, novel wavelength-tunable optoelectronic devices would be viable in BLG.

\begin{figure}[p]
\begin{center}
\includegraphics[width=\figwidth]{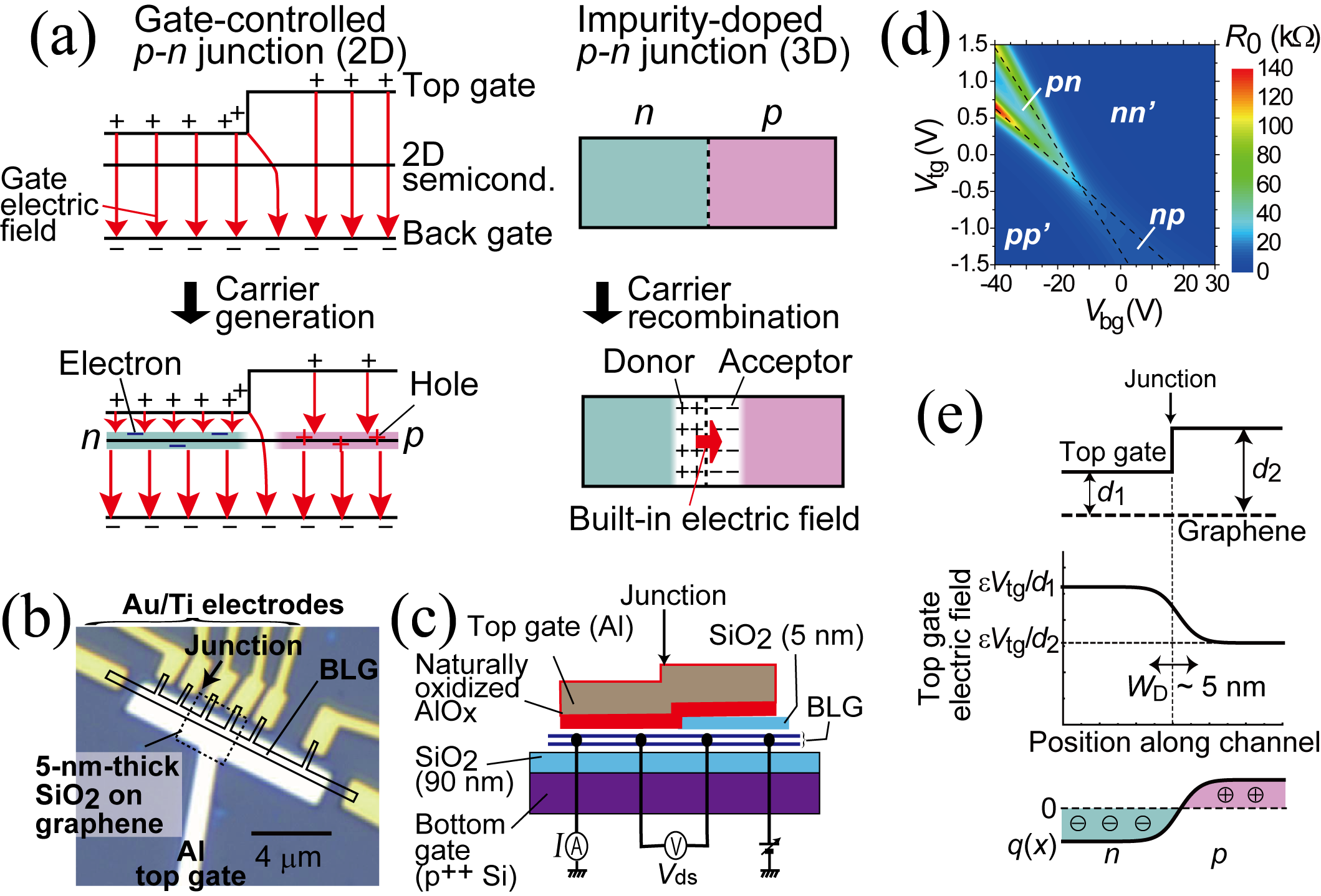}%
\end{center}
\caption{(Color online) (a) Schematic diagram of \pn junction mechanism for a gate-controlled 2D semiconductor (left) and an impurity-doped 3D semiconductor (right). (b) Optical micrograph of a BLG \pn junction with a stepwise top gate. The solid lines show the shape of the graphene channel made by oxygen plasma etching. The dashed lines show the area in which the graphene is covered by a 5-nm-thick SiO$_2$ layer. 
(c) Schematic illustration of the cross-sectional view of the \pn junction. The locally inserted SiO$_2$ layer creates the stepwise structure in the top gate. The instrumental configuration for a four-terminal measurement is also shown. All transport properties shown in this paper were acquired in the same configuration. (d) Color map of the linear resistance, $R_0$, as a function of \Vbg and \Vtg at $T=80$ K, determined at $V_{\rm ds}\sim 1$ mV. The dashed lines show the ridges of the CNPs separating $p$ and $n$ regions.  (e) Schematic view of the geometrical structure of the stepwise top gate (top), the profile of the top gate electric field (middle), and the charge distribution (bottom). The top gate electric field is estimated numerically (Appendix A).}
\end{figure}

\begin{figure}[p]
\begin{center}
\includegraphics[width=\figwidth]{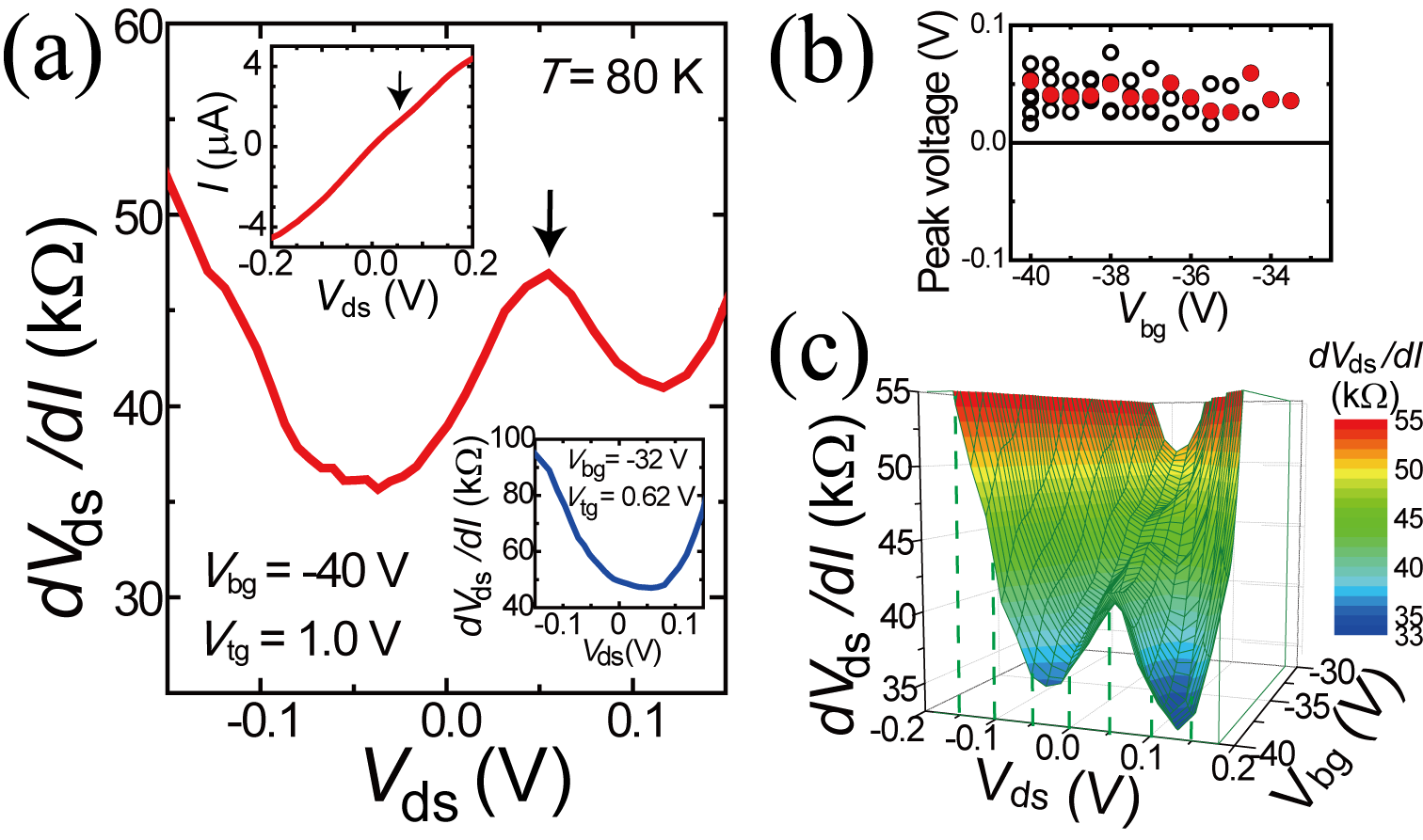}%
\end{center}
\caption{(Color online) (a) Differential resistance, \dVdI, as a function of the source-drain voltage \Vds for BLG under a relatively large electric field ($V_{\rm bg} = -40$ V). The arrow indicates the \dVdI peak. 
The top left inset shows current-voltage curve corresponding to the main panel.
The bottom-right inset shows a $dV_{\rm ds}/dI$-$V_{\rm ds}$ curve for a relatively small electric field($V_{\rm bg} = -32$ V). 
(b) Distribution of the $V_{\rm ds}$ values for the $dV_{\rm ds}/dI$ peaks as a function of $V_{\rm bg}$ with various values of $V_{\rm tg}$. The red closed symbols correspond to the data shown in (c). (c) $dV_{\rm ds}/dI$ as a function of \Vds and \Vbg. The \Vtg value is selected along the middle point of two ridges of CNPs in Fig.\ 1(c). All data were acquired at $T=80$ K.}
\end{figure}

\begin{figure}[p]
\begin{center}
\includegraphics[width=\figwidth]{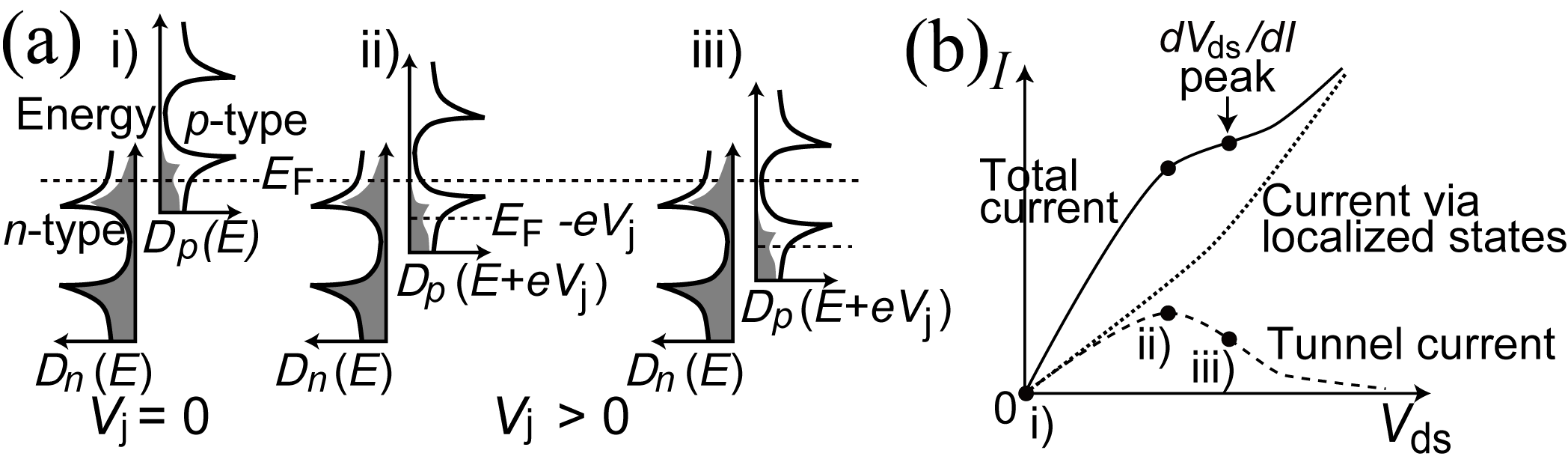}%
\end{center}
\caption{(a) Schematic diagram of the DOS for the $p$-type and $n$-type sides: i) for the unbiased condition ($V_{\rm j} = 0$) and ii)-iii) for the forward-biased ($V_{\rm j} > 0$) conditions. The gray shaded areas show the density of occupied states with a Fermi distribution. Electron tunneling takes place from occupied states on one side to unoccupied states on the other side. (b) Schematic $I$-$V_{\rm ds}$ curve for the tunneling current (dashed curve), the current via localized states (dotted curve), and the total current (solid curve). The three conditions in (a) ( i), ii), and iii) ) are identified on the tunneling current curve and the total current curve.}
\end{figure}

\begin{figure}[p]
\begin{center}
\includegraphics[width=\figwidth]{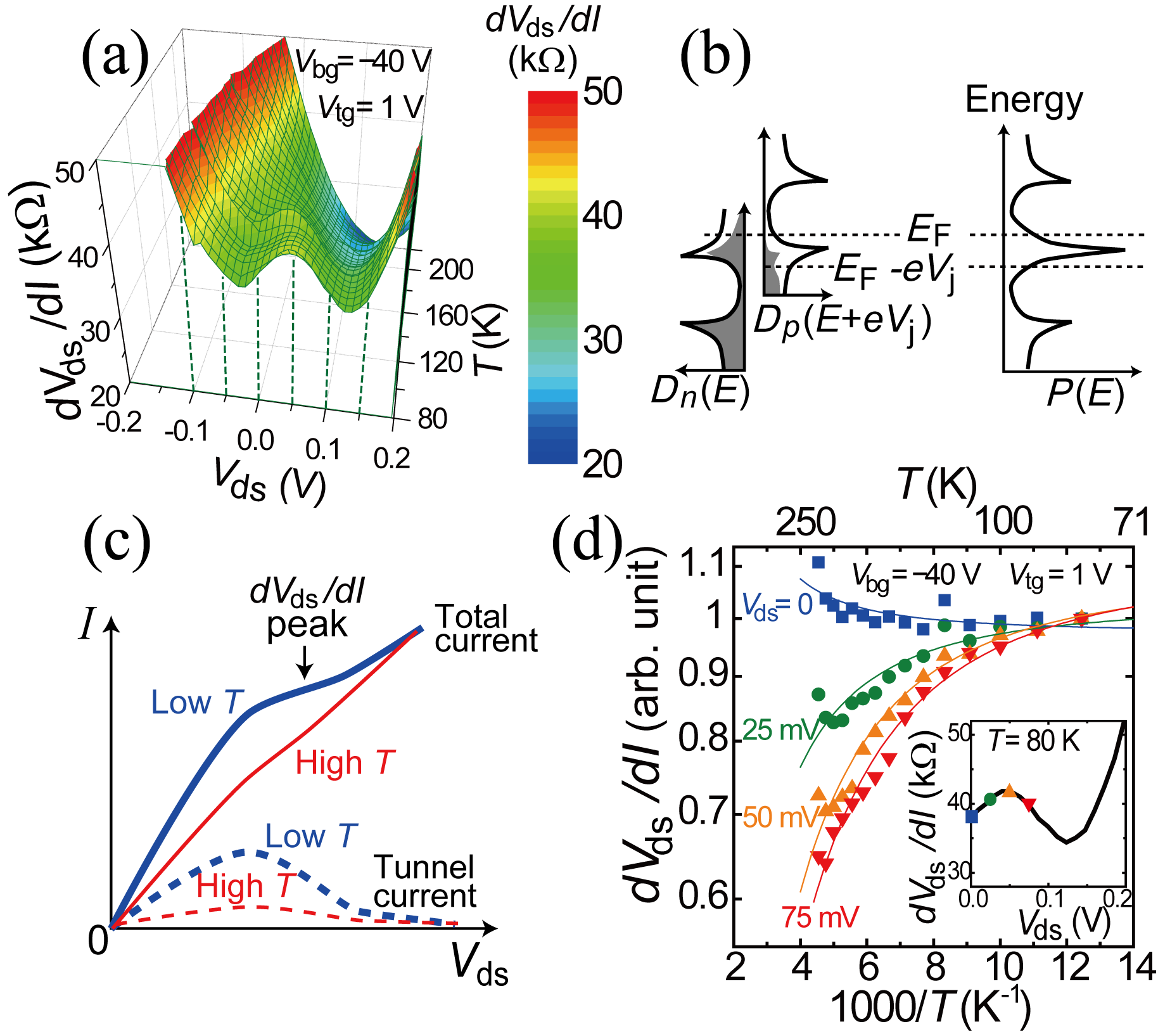}%
\end{center}
\caption{(Color online) (a) Temperature dependence of the \dVdI curve for $V_{\rm bg} = -40$ V and $V_{\rm tg} = 1.0$ V. (b) Schematic diagram of $A(E) = D_n(E)D_p(E + eV_{\rm j})$ under a forward voltage drop of \Vj corresponding to the tunneling current peak (Fig.\ 3(a)ii)). (c) Schematic current-voltage characteristics for the tunneling current at low temperature (thick dashed curve) and high temperature (thin dashed curve) and the total current with an additional leakage current (Fig.\ 3(b)) at low temperature (thick solid curve) and high temperature (thin solid curve). (d) Arrhenius plot of the \dVdI values extracted from a at fixed values of \Vds for a forward bias ($V_{\rm ds} > 0$). The \dVdI values are normalized at $T = 80$ K. The solid lines represent fittings within second order of the temperature. The \dVdI curve is shown for a forward bias at $T = 80$ K in the inset of (b), in which the markers indicate the corresponding \Vds values in the main panel. }
\end{figure}

\begin{acknowledgments}
This work was supported in part by KAKENHI (No. 21241038) from the MEXT of Japan and by the FIRST Program from the JSPS. The authors would like to thank S. Okada, M. Otani, M. Koshino, and K. Wakabayashi for useful discussions. We also thank Covalent Materials Corporation for providing Kish graphite as the source material of graphene. 
\end{acknowledgments}

\renewcommand{\thefigure}{
\thesection.\arabic{figure}}
\setcounter{figure}{0}

\appendix
\section{Width of the Depletion Region}

We derive a Poisson equation to describe the electric potential profile in a double-gated 2D semiconductor.
Here, we define a 2D conductor as a thin conductor which is much thinner than electric field screening length.
Monolayer and bilayer graphene fit to the definition \cite{miyazakiAPEXscr}.
In the 2D semiconductor, the charge distribution along the thickness direction is negligible.
The 2D semiconductor (thickness $d_{\rm s}$, dielectric constant $\varepsilon_{\rm s}$) is sandwiched between a bottom gate and a top gate (Fig.\ A$\cdot$1(a)).
We consider a small region in the semiconductor from a position $x$ to a position $x+\Delta x$. 
Gauss's law gives a relationship among the top gate electric field ($E_{\rm t}$), the bottom gate electric field ($E_{\rm b}$), the electric field in the semiconductor plane ($E_{\rm i}$), and charge density in the small region $q(x)$: 
$\varepsilon_{\rm s}\left( E_{\rm i}(x+\Delta x)- E_{\rm i}(x) \right)d_{\rm s}
+(\varepsilon_{\rm t} E_{\rm t}-\varepsilon_{\rm b} E_{\rm b})\Delta x=q(x)\Delta x$, with $\varepsilon_{\rm t}$ and $\varepsilon_{\rm b}$ as the dielectric constant of the top gate insulator and the bottom gate insulator, respectively. 
Taking the limit as $\Delta x$ approaches zero, we have a one-dimensional Poisson equation for the electric potential ($\it \Psi$) in the 2D semiconductor, 
\begin{equation}
 \frac{d^2 \Psi}{dx^2}=-\frac{dE_{\rm i}}{dx}= \frac{\varepsilon_{\rm t} E_{\rm t}-\varepsilon_{\rm b} E_{\rm b}-q(x)}{\varepsilon_{\rm s}d_{\rm s}}. 
\end{equation}

Here, we assume that the top gate electric field is changed abruptly at the junction, such as $E_{\rm t}=E_{\rm t \it n}$ for $x\leq 0$ and $E_{\rm t}=E_{\rm t \it p}$ for $x > 0$ (abrupt junction).
The $\it \Psi$ is constant in homogeneous regions which are sufficiently distant from $x=0$.
In these regions, the electric field in the semiconductor is zero ($E_{\rm i} = -d{\it \Psi}/dx = 0 $), and the charge density is homogeneous ($q(x)= \varepsilon_{\rm t} E_{\rm t}-\varepsilon_{\rm b} E_{\rm b}$).
A $p$-$n$ junction is formed at $x=0$ for $eN_p\equiv \varepsilon_{\rm t} E_{\rm t \it p}-\varepsilon_{\rm b} E_{\rm b}>0$ and $-eN_n\equiv \varepsilon_{\rm t} E_{\rm t \it p}-\varepsilon_{\rm b} E_{\rm b}< 0$.
There is a transient region form $p$ to $n$ around $x=0$.
If the 2D conductor has the band gap, charge carriers are depleted from the transient region. 
Assuming that $|q(x)|$ is much smaller than $eN_p$ and $eN_n$, the Poisson equation in the depletion region becomes,
\begin{eqnarray*}
\frac{d^2 {\it \Psi}}{dx^2} &=& \frac{eN_p}{\varepsilon_{\rm s}d_{\rm s}} \quad \mbox{(for the $p$ side)},\\
-\frac{d^2 {\it \Psi}}{dx^2} &=& \frac{eN_n}{\varepsilon_{\rm s}d_{\rm s}} \quad \mbox{(for the $n$ side)}.
\end{eqnarray*}
These equations have same form as in a 3D $p$-$n$ junction\cite{sze}, in which the acceptor density ($N_{\rm A}$) and the the donor density ($N_{\rm D}$) are replaced by $N_p/d_{\rm s}$ and $N_n/d_{\rm s}$, respectively.
Use the same solution method for the 3D case, we obtain a depletion region width for the abrupt junction,
\[ W_{\rm D}^{\rm abrupt} = \sqrt{ \frac{2\varepsilon_{\rm s}d_{\rm s}}{e} \frac{N_p+N_n}{N_pN_n} \Delta{\it \Psi}}, \]
where $\Delta{\it \Psi}$ is the electrical potential difference between the $p$ side and the $n$ side.
Using typical values in our experiment ($N_p\sim N_n\sim 2\times 10^{12}\ {\rm cm}^{-2}$, $\varepsilon_{\rm s} = 3.0\varepsilon_{\rm 0}$ for graphite\cite{DresselhausCNT}, $d_{\rm s}=0.67$ nm for bilayer graphene, and $\Delta{\it \Psi} \sim E_{\rm g}=0.2$ eV), the transient region width is $W_{\rm D}^{\rm abrupt}\sim 2$ nm.

The abrupt junction model is applicable only when the transient region, in which the top gate electric field changes from $E_{\rm t \it n}$ to $E_{\rm t \it p}$, is narrower than the $W_{\rm D}^{\rm abrupt}$.
If not, we have to take the transient region width as the depletion region width $W_{\rm D}$\cite{sze}.
In the stepwise top gate, the top-gate dielectric thickness is changed from $d_1$ to $d_2$ ($d_1<d_2$) at position $x=0$ (thick solid line in Fig.\ A$\cdot$2(a)).
An electric potential, $\phi(x,z)$, between graphene and the top gate is obtained by solving a Poisson equation numerically with a boundary condition, $\phi=0$ at the graphene ($z=0$) and $\phi=1$ at the top gate ($z=d_1$ for $x\leq 0$, and $z=d_2$ for $x>0$) (Fig.\ A$\cdot$2(a)).
In the calculation, SiO$_2$-equivalent thickness of the dielectric layer ($d_1=3.7$ nm and $d_2=6.3$ nm) is used.
The electric field component normal to the graphene, $E_\perp\propto -{\partial \phi}/{\partial z}|_{z=0}$, changes with a 5-nm-wide transient region around $x=0$ (Fig.\ A$\cdot$2(b)).
In the transient region, the electric field has a tangent component to the graphene, $E_\parallel \propto -{\partial \phi}/{\partial x}$, at the vicinity of the graphene surface (Fig.\ A$\cdot$2(c)).
Since the transient region is wider than the $W_{\rm D}^{\rm abrupt}$($\sim 2$ nm), the transient region width gives the depletion region width, i.e., $W_{\rm D}\sim 5$ nm.

\begin{figure}[p]
\begin{center}
  \includegraphics[width=130mm]{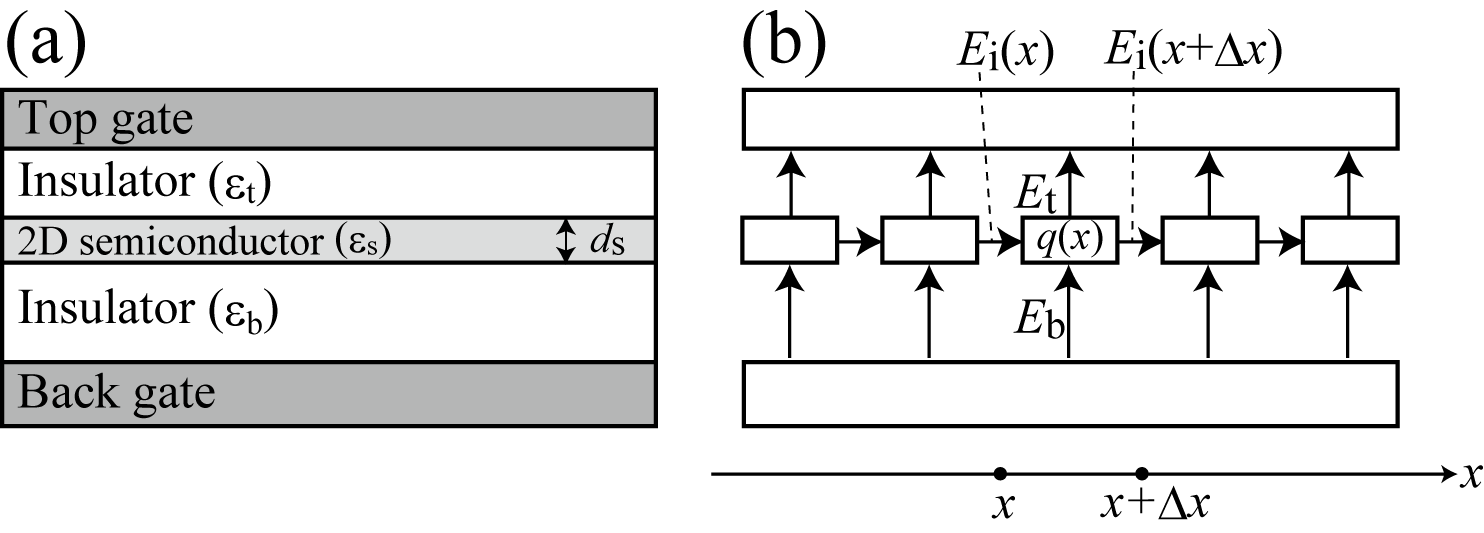}
 \end{center}
\caption{Schematic illustrations of (a) 2D semiconductor sandwiched between a top gate and a bottom gate and (b) electric field in the semiconductor plane ($E_{\rm i}$) and in gate insulators ($E_{\rm t}$ and $E_{\rm b}$)}.  
\end{figure}

\begin{figure}[p]
\begin{center}
  \includegraphics[width=130mm]{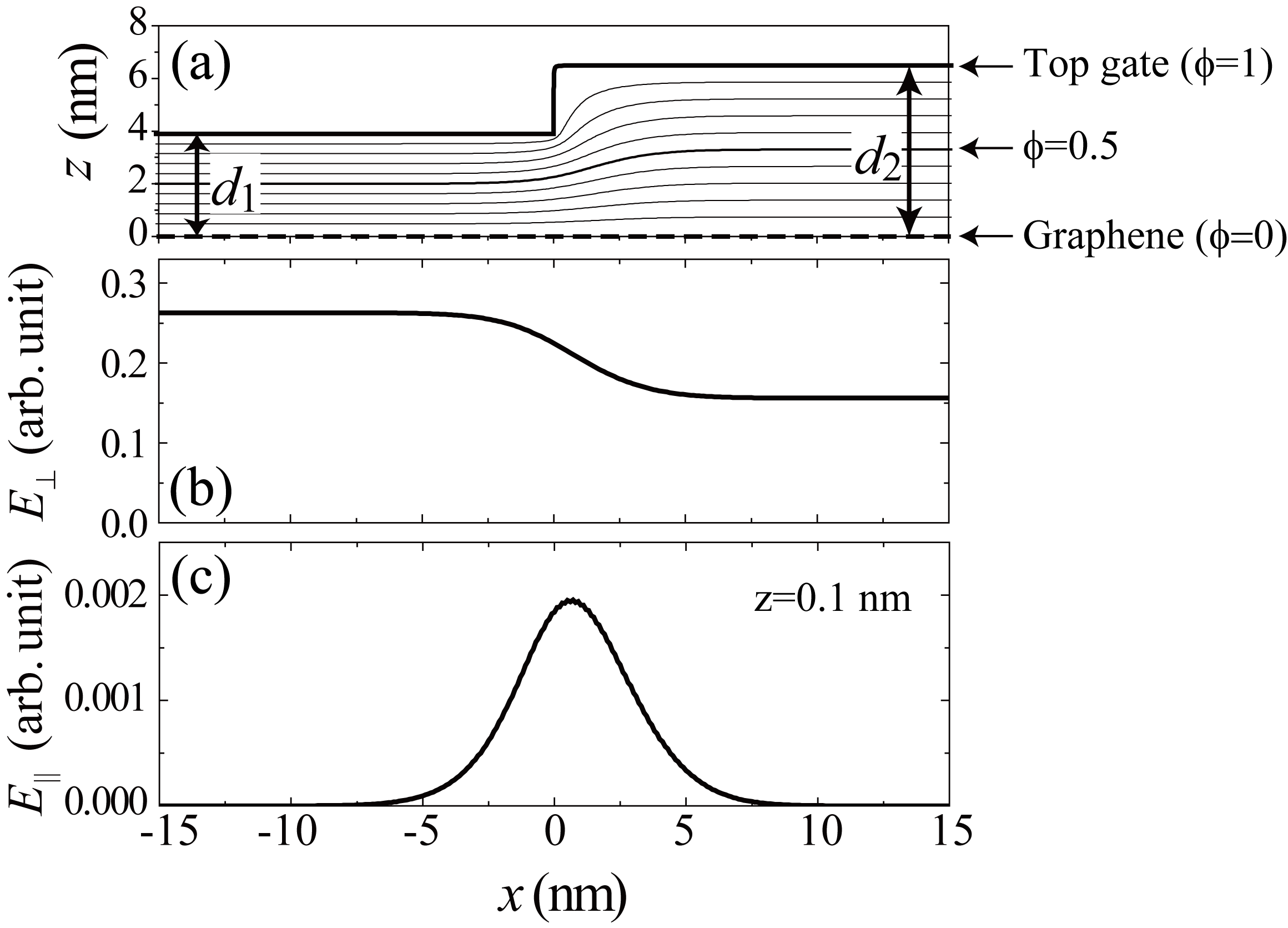}
 \end{center}
\caption{(a) Contour plot of electric potential $\phi$ with a boundary condition that $\phi=0$ at the graphene and $\phi=1$ at the stepwise top gate. The contour interval is 0.1. (b) Normal component of the electric field at the graphene surface. (c) Tangent component of the electric field at the vicinity of the graphene surface ($z=0.1$ nm).}
\end{figure}

\section{Estimation of Band Diagram and Junction Resistance}
Equation (A$\cdot$1) gives the charge density $q(x)$ in graphene sandwiched by top and bottom gates, $ q(x) = \varepsilon_{\rm t} E_{\rm t}-\varepsilon_{\rm b} E_{\rm b}+\varepsilon_{\rm s}d_{\rm s} dE_{\rm i}/dx $.
The gate electric fields are proportional the gate voltages, giving $q(x)=-\Ctg (x) V_{\rm tg} - \Cbg V_{\rm bg} +\varepsilon_{\rm s}d_{\rm s} dE_{\rm i}/dx +q_0 $ with $C_{\rm tg(bg)}$ as local capacitance for the top (bottom) gate and $q_0$ as the charge density for $V_{\rm bg}=V_{\rm tg}=0$.
Three typical band diagrams for graphene with spatially modulated charge density are mentioned in Fig.\ B$\cdot$1(a)-1(c), corresponding to $p$-$p$, $n$-$p$, and $n$-$n$ combinations. 
In the resistance measurement, these combinations are separated by the two ridges of CNP (Fig.\ B$\cdot$1(d)).
On the CNPs, the Fermi level is at the mid-gap of the right or left side in the stepwise modulation.

For the \Vtg fixed in between the two CNPs, a spatial modulation forming a junction with $p$- and $n$-region is generated.
When the \Vbg is applied to the \pn junction, the $p$- and the $n$-region are separated by the tunneling barrier caused by the band gap (Fig.\ B$\cdot$1(b)).
The charge density in each $p$- (or $n$-) region is given by $q_{p({\rm or}\; n)}=-\Ctg(V_{\rm tg}-V_{\rm CN}^{p({\rm or}\; n)})$, where $V_{\rm CN}^{p({\rm or}\; n)}$ is the top gate voltage for the CNP in the $p({\rm or}\; n)$-region.
When the $V_{\rm bg}=-40$ V and $V_{\rm tg}=1.0$ V (corresponding to ``(b)'' in Fig.\ B$\cdot$1(d)) are applied, $q_p/e=1.7\times 10^{12}\ {\rm cm^{-2}}$ (or $q_n/e=2.5\times 10^{12}\ {\rm cm^{-2}}$) is extracted from $V_{\rm CN}^p=1.5$ V (or $V_{\rm CN}^n=0.57$ V).
The carrier density $q_{p({\rm or}\; n)}/e$ equals to integration of the density of states (DOS) from the mid-gap to the Fermi level. 
The typical charge density $q_p/e \sim q_n/e \sim 2\times 10^{12}\ {\rm cm}^{-2}$ in our experiment is similar to that in reported experiments on graphene \pn junction \cite{HuardPRLpnp, GorbachevNLpnp}.

An ideal DOS ($D$) without band tail is given as a function of the energy ($\varepsilon$) measured from the mid-gap: $D(\varepsilon) \sim (t_\perp/c^2) \sqrt{\Delta/|\varepsilon-E_{\rm g}/2|}$ for $|\varepsilon|> E_{\rm g}/2$ and $D(\varepsilon)=0$ for $|\varepsilon| \leq E_{\rm g}/2$, with $\Delta$ as the potential difference between the two graphene layers and $c(\sim 1\times 10^6\ {\rm m/s})$ is the Dirac velocity in graphene \cite{mkhiraryan}.
The $\Delta$ is related to the potential difference between the two graphene layers by $E_{\rm g} = 2(\Delta-\Delta^3/{t_\perp}^2)$, where $t_\perp$($\sim 0.3$ eV) is the interlayer hopping integral\cite{GuineaPRB73}, giving $\Delta\sim 0.12$ eV for $E_{\rm g}\sim 0.2$ eV. Because the DOS of the graphene has a singularity at the band edges ($\varepsilon=\pm E_{\rm g}/2$) \cite{mkhiraryan, nilsson}, the Fermi level $E_p$ (or $E_n$) measured from the band edge of the $p$- (or $n$-) region is related to the charge density $q_p$ (or $q_n$) as $q_i/e = -\int_{0}^{\alpha_i (E_{\rm g}/2+E_i)} D(\varepsilon) d\varepsilon = -\alpha_i(2t_\perp/{c}^2)\sqrt{\Delta\cdot E_i}$ ($i=p$ or $n$), where $\alpha_p=-1$ and $\alpha_n=1$.
Using typical values of the carrier density of $q_p/e=q_n/e\sim 2\times 10^{12}\ {\rm cm^{-2}}$ with potential difference between the two layers $\Delta\sim 0.12$ eV and $E_{\rm g}=0.2$ eV in our experiment, the Fermi level of the $E_p$ and $E_n$ in the each side of $p$-$n$ junction are a few meV.
These values are two orders smaller than the energy gap $E_{\rm g}$, indicating that the $E_p$ and the $E_n$ are very close to the singularity peak in the each regions.
In a realistic semiconductor graphene with band tails caused by localized states, the singularity peak becomes broader with increasing the density of localized states \cite{mkhiraryan, nilsson}.
Except for the graphene extremely disordered, the $E_p$ and $E_n$ are still much smaller than the $E_{\rm g}$, and are close to the band edges.
Thus, the band diagram measured in the $p$-$n$ junction can be described as Fig.\ B$\cdot$1(b).

The magnitude of the junction resistance can be roughly estimated from the band diagram. The junction resistance consists of a parallel combination of a tunnel component and a leakage component. The tunnel component is estimated by the tunneling probability of an electron at the junction and the number of states involved in the tunneling. At the junction, the potential gradient is $F\sim E_{\rm g}/W_{\rm D}\sim 0.04$ eV/nm, which leads to a tunneling probability across the junction of $T_{\rm t}=\exp\left[-4\sqrt{2m^*}{E_{\rm g}}^{3/2}/(3\hbar F)\right]\sim 0.07$, where $m^*={t_\perp}^2/4\Delta c^2 \sim 0.03 m_{\rm e}$ for $\Delta = 0.12$ eV, $m_{\rm e}$ is the bare electron mass, and $c\sim 1\times 10^6$  m/s is the Dirac velocity in graphene, for low energy states in a ``Mexican hat'' band structure\cite{sze, mkhiraryan, mccann, GuineaPRB73}. Landauer's formula gives the tunnel conductance $1/R_{\rm t}\sim N(e^2/\pi\hbar)T_{\rm t}$ , where $N$ is the number of quantized wavenumber states in the channel-width direction and $\hbar$ is Planck's constant. Even at the band edge, the Mexican hat band structure has a non-zero wavenumber range from 0 to $k_0$, where $k_0=\Delta/\hbar c$ ($\sim 0.2$ nm$^{-1}$ for $\Delta= 0.12$ eV)\cite{mccann,GuineaPRB73}. The range of the wavenumber in the $p$-type and $n$-type region is approximately 0 to $k_0$, because the Fermi level is close to the band edge. $N$ is roughly $k_0W/\pi \sim 30$ ($W \sim 0.4$ $\mu$m is the channel width) because the wavenumber is quantized\cite{WeesQPT} by $\pi/W$. Thus, the tunnel resistance $R_{\rm t}$ is estimated to be several k$\Omega$. The leakage conduction via localized states coexists with the tunnel conduction. The leakage resistance is expressed as {\bf $R_{\rm L}\sim \rho_0 W_{\rm D}$}, where $\rho_0$ is the residual resistance at the mid-gap states; that is, the resistance per unit length at the CNP. Using $\rho_0\sim 0.2$ k$\Omega$/nm for $V_{\rm bg} = -40$ V, the leakage resistance is on the order of $R_{\rm L} \sim 1$ k$\Omega$. Thus, the leakage current is comparable to the tunneling current. Finally, the junction resistance obtained, $R_{\rm j} = 1/(1/R_{\rm t} + 1/R_{\rm L})$, is on the order of 1 k$\Omega$. Using this value, it is estimated that the voltage drop at the junction \Vj is a few percent of $V_{\rm ds}$, corresponding to a few mV, because the total resistance of the sample is about 40 k$\Omega$.

\begin{figure}[p]
\begin{center}
  \includegraphics[width=130mm]{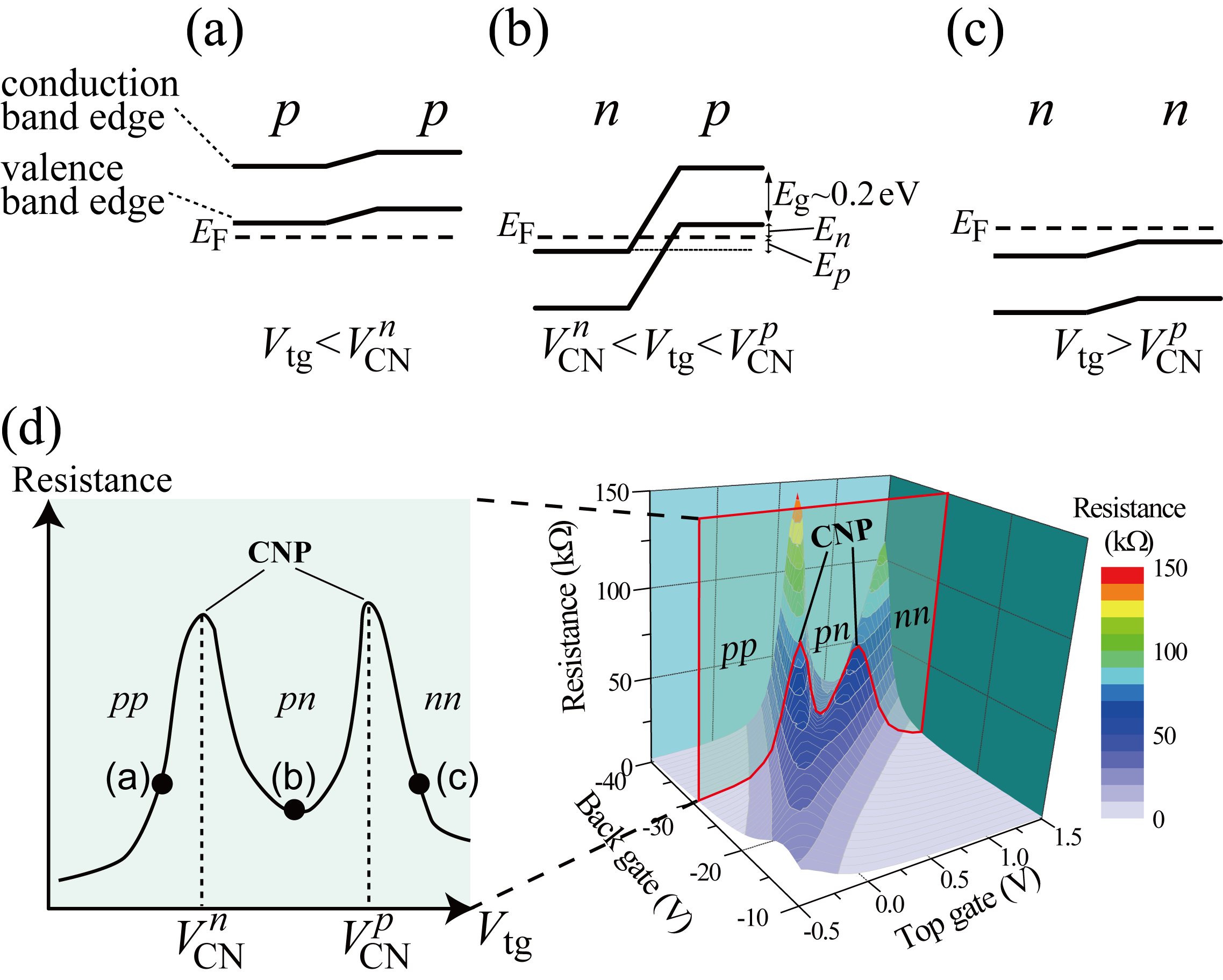}
 \end{center}
\caption{(Color online) Schematic band diagrams for combinations of (a) $p$-$p$, (b) $n$-$p$, and (c) $n$-$n$ in the semiconductor graphene with the stepwise top gate. 
(d) The left panel shows resistance change as a function of the top gate voltage at fixed \Vbg, and the right panel shows a 3D plot of the resistance as a function of the bottom gate and top gate voltage for the same data in the middle panel of Fig.\ 1(c).
The left panel corresponds to the cross section in the right panel. 
The marked three points in the right panel correspond to the band diagrams (a)-(c).}
\end{figure}

\bibliographystyle{jpsj}
\bibliography{pnbib}

\end{document}